\documentclass[a4paper]{article}

\usepackage[english]{babel}
\usepackage[T1]{fontenc}
\usepackage[utf8]{inputenc}
\usepackage{textcomp}
\usepackage[a4paper,top=3cm,bottom=2cm,left=3cm,right=3cm,marginparwidth=1.75cm]{geometry}

\usepackage{amsmath}
\usepackage{graphicx}
\usepackage[colorinlistoftodos]{todonotes}
\usepackage[colorlinks=true, allcolors=blue]{hyperref}
\usepackage{booktabs}
\usepackage{graphicx}
\usepackage{hyperref}
\usepackage{listings}
\usepackage{amsmath}
\usepackage{verbatim}
\usepackage{color}
\usepackage{wrapfig}
\usepackage{lineno}
\usepackage{array}
\usepackage{makecell}
\usepackage{tabularx}
\newcolumntype{Y}{>{\centering\arraybackslash}X}
\usepackage{subcaption}
\usepackage{caption}
\usepackage{listings}
\usepackage{xcolor}
\usepackage{colortbl}
\usepackage{hyperref}
\usepackage{authblk}
\usepackage{longtable}
\usepackage{adjustbox}
\usepackage{multirow}
\usepackage{float}
\usepackage{setspace}
\usepackage{pdfpages}
\usepackage[style=nature,url=true,doi=true,maxbibnames=3, defernumbers=true]{biblatex}
\bibliography{sample.bib}
\usepackage{pdflscape}
\newcolumntype{L}[1]{>{\raggedright\arraybackslash}p{#1}}
\newcolumntype{Y}{>{\raggedright\arraybackslash}X}

\author[1]{Waqar Muhammad Ashraf}
\author[1]{Diane Coyle}
\author[1*]{Ramit Debnath}

\date{} 

\title{Code, Capital, and Clusters: Understanding Firm Performance in the UK AI Economy}

\affil[1]{University of Cambridge, Cambridge, CB2 1PZ, UK}
\affil[*]{corresponding author: rd545@cam.ac.uk}

\begin{document}
\maketitle

\begin{abstract}
The UK has established a distinctive position in the global AI landscape, driven by rapid firm formation and strategic investment. However, the interplay between AI specialisation, local socio-economic conditions, and firm performance remains under-explored. This study analyses a comprehensive dataset of UK AI entities (2000–2024) from Companies House, ONS, and glass.ai. We find a strong geographical concentration in London (41.3\% of entities) and technology-centric sectors, with general financial services reporting the highest mean operating revenue (£33.9 million, n=33). Firm size and AI specialisation intensity are primary revenue drivers, while local factors—Level 3 qualification rates, population density, and employment levels—provide significant marginal contributions, highlighting the dependence of AI growth on regional socio-economic ecosystems. The forecasting models project sectoral expansion to 2030, estimating 4,651 [4,323–4,979, 95\% CI] total entities and a rising dissolution ratio (2.21\% [-0.17–4.60\%]), indicating a transition toward slower sector expansion and consolidation. These results provide robust evidence for place-sensitive policy interventions: cultivating regional AI capabilities beyond London to mitigate systemic risks; distinguishing between support for scaling (addressing capital gaps) and deepening technical specialisation; and strategically shaping ecosystem consolidation. Targeted actions are essential to foster both aggregate AI growth and balanced regional development, transforming consolidation into sustained competitive advantage.
\end{abstract}

\section{Introduction}

Artificial intelligence (AI) has matured from a theoretical pursuit into the defining general-purpose technology of the Fourth Industrial Revolution \cite{skilton20184th}. Although early software firms focused on digitising business operations \cite{culler1999parallel}, the trajectory of the sector shifted dramatically after breakthroughs in deep learning and hardware capabilities in 2012 \cite{hinton2006fast, krizhevsky2012imagenet}. These advances enabled the transition from theoretical algorithms to commercially viable products, driving significant economic activity \cite{brynjolfsson2017artificial}. Today, the global landscape is characterised by a surge in AI-related entities, ranging from university spinouts to large-scale incubators, fuelled by the generative AI boom and the widespread commercialisation of Large Language Models (LLMs) \cite{babina2024artificial, baraybar2025ai}.

In this global context, the United Kingdom (UK) has established itself as a critical player, currently ranked third globally behind only the USA and China with a sector valuation of £72.3 billion \cite{businessgov2025grow}. The UK ecosystem is robust: in 2024 alone, the number of operating AI entities increased by 58\% to 5,862, generating an estimated £23.9 billion in revenue \cite{dsit2025sector}. Crucially, this revenue is highly concentrated; large firms, constituting just 5\% of registered entities, generate 85\% of total revenue (£20.4 billion). The government has adopted a pro-innovation policy regime with statements including the ``Artificial Intelligence Action Plan" and the ``Industrial Strategy`` pledging significant investment in academic and physical infrastructure to support AI growth \cite{dsit2025actionplan, ukgov2023boost, ukgov2025indstrategy}. However, while aggregate statistics highlight the sector's scale, they often obscure the structural dynamics of value creation. For instance, the distinction between ``dedicated" AI firms (developing core AI IP) and ``diversified" entities (integrating AI into existing tech stacks) remains critical, yet under-analysed in relation to firm survival and revenue efficiency.

Despite the wealth of high-level government reports that identify capital and broad talent pools as key drivers \cite{technation2024sector, bone2025skills}, there remains a notable absence of granular, firm-level analysis. The current literature establishes macro-growth drivers but lacks a detailed understanding of how micro factors, such as specific technical specialisations (e.g., natural language processing vs. machine vision) and hyper-local socio-economic conditions, influence firm performance \cite{raj2018artificial, roper2024understanding}. Specifically, the link between postcode-level demographics (such as local qualification levels and professional density) and the operating revenue of locally registered AI firms is largely unmapped. Furthermore, as the sector matures, policymakers require predictive insights into the lifecycle of these firms: identifying not just what drives formation but what predicts dissolution or sustainable scaling in a competitive market.

To address these gaps, this article presents a data-driven mapping study of the UK AI landscape. We analyse a comprehensive longitudinal dataset of firm-level data between 2000 and 2024, compiled from public administrative sources including UK Companies House, the Office of National Statistics (ONS), and glass.ai. By augmenting firm registries with postcode-level socio-economic indicators, we construct a multidimensional view of the sector. We employ CatBoost gradient boosting models embedded within a SHapley Additive exPlanations (SHAP) framework to isolate the specific impact of technical specialisation and local geography on operating revenue. Finally, we forecast sector trajectories up to 2030, modelling active versus dissolved entity ratios. These findings provide empirical data-driven evidence to guide targeted policy interventions, moving beyond aggregate growth metrics to support a balanced and resilient AI economy.

\section{Methods}

\subsection{Data collection}

The primary dataset for this study was constructed by integrating administrative records from government repositories with web-crawled firm data. We established a baseline using the UK Research and Innovation (UKRI) "WAIFinder" database, which catalogues 3,280 AI-related organisations \cite{ukri2024waifinder}. This repository provides foundational attributes for each entity, including registered name, technical specialisation (e.g., machine vision, NLP, big data), entity type (e.g., company, university, incubator) and geolocation data (postcode and coordinates).

To ensure that the study focusses on the primary creators of AI value, we applied strict inclusion criteria. We retained only entities classified as ``companies" or ``universities", organisations directly involved in developing AI products, services, and tangible infrastructure. Facilitators of the ecosystem, such as incubators, accelerators, and funding bodies, were excluded from the analysis because they typically do not develop proprietary AI technology for business deployment.

This baseline was subsequently updated and expanded to cover the period up to 2024. We used the semantic web-reading capabilities of glass.ai \cite{glassai2025} to identify new market entrants and cross-referenced these records with Companies House \cite{companieshouse2025} and LinkedIn \cite{linkedin2025} to verify the years of registration and operational status. During this harmonisation process, we simplified the diverse range of operational statuses reported by Companies House. Entities listed as ``liquidation," ``administration," ``removed," or ``voluntary arrangement" were mapped to a single ``Dissolved" category, while all others were classified as ``Active." Following this filtration and augmentation process, the final dataset comprised 4,392 unique AI entities.

Finally, to investigate the impact of local ecosystem characteristics on firm performance, we enriched the dataset with socio-economic indicators at the postcode-level of the Office of National Statistics (ONS) 2021 Census \cite{ons2025census}. For each registered entity location, we extracted data on workforce composition and human capital, specifically: population density, the proportion of residents with Level 3 (equivalent to A-Levels) and Level 4+ (high education) qualifications, and employment density in key sectors (Information and Communication; Financial and Insurance; Professional Occupations). These variables allow us to model the association between local socio-economic capital and the survival and revenue trajectories of AI firms.  

\subsection{Data processing}

We extracted operating revenue and workforce size (employee count) data for all entities where available. Revenue figures were sourced from Companies House filings and publicly available annual reports. However, data availability is constrained by UK financial reporting standards, which allow "small companies" (defined as having a turnover under £10.2m or fewer than 50 employees) to file abbreviated accounts that exclude profit and loss statements. Given that the AI sector is dominated by early-stage ventures and small-and-medium enterprises (SMEs), a significant proportion of the dataset utilizes these reporting exemptions.

To mitigate data sparsity with respect to the size of the workforce, we triangulated Companies House records with LinkedIn employment data. Despite these efforts, the requirement for complete financial disclosure limits revenue-specific analysis to a subset of 529 entities that publicly reported in their accounting statements operating revenues and employee counts.

To quantify the technical focus of each organisation, we analysed the working domains and descriptors listed on entity websites. We applied Term Frequency-Inverse Document Frequency (TF-IDF) to convert qualitative descriptions of AI specialisation (e.g., ``computer vision," ``predictive analytics") into quantitative keyword scores. TF-IDF assigns weight to each term, reflecting its importance within a specific document relative to the entire corpus. For each entity, we calculated a composite specialisation score by aggregating the TF-IDF weights of its associated keywords. This provides a numerical metric that represents the intensity and distinctiveness of the entity's technical portfolio.

\subsection{Model development and feature importance}

Prior to modelling, we processed both categorical and continuous variables to ensure compatibility with the learning algorithm. Categorical predictors, specifically city and sector group, were one-shot encoded to transform nominal data into a binary vector format. For the target variables, \texttt{operating revenue} and \texttt{operating revenue per employee}, we applied a logarithmic transformation. This was necessary to normalize the data distribution, mitigate the impact of positive skewness (long-tailed distributions), and stabilize variance across entities with vastly different operating scales.

We employed the CatBoost (Categorical Boosting) algorithm to model the relationship between firm-level and postcode-level input variables and the financial performance metrics. CatBoost is a gradient boosting on decision trees (GBDT) framework chosen for its state-of-the-art performance on tabular datasets, often outperforming traditional ensemble methods and neural networks in this domain \cite{ramakrishnan2025neural}.

Two distinct models were trained to predict \texttt{operating revenue} and \texttt{operating revenue per employee}, respectively. To ensure robust generalisation and minimize overfitting, we conducted rigorous hyperparameter tuning via an iterative optimisation approach. The predictive accuracy and reliability of the trained models were assessed using the following metrics:

\begin{equation}
R^2 = 1 - \frac{\sum_{i=1}^N (y_i - \hat{y}_i)^2}{\sum_{i=1}^N (y_i - \bar{y_i})^2}
\end{equation}
\begin{equation}
RMSE = \sqrt\frac{\sum_{i=1}^N (y_i - \bar{y_i})^2}{N}
\end{equation}

here, $y_i$ is the actual while $\hat{y}_i$ is the model-based response for observation, i = 1,2,3, …N. whereas $\hat{y}_i$ is the mean calculated in $y_i$. R$^{2}$ is considered as an accuracy measure, and it varies from zero to one. The R$^{2}$ score computed in the data set and approaching one may indicate a good map between the actual and model-based responses. However, error metrics such as RMSE should also be computed to estimate the aggregated mean error in the model-based predictions. CatBoost models are trained on the data partition in to training (80\%) and testing (20\%) datasets and the modelling performance is computed on the two metrics.

To interpret the predictive outputs of the CatBoost models, we employed the SHapley Additive exPlanations (SHAP) framework. SHAP is grounded in cooperative game theory and calculates the marginal contribution of each input variable toward the final prediction \cite{lundberg2017unified}. Widely adopted for quantifying feature significance in regression tasks \cite{futagami2021pairwise}, this framework allows us to estimate the relative contribution of each feature. Consequently, we utilise SHAP to identify the key variables driving \texttt{operating revenue} and \texttt{operating revenue per employee} for AI entities operating within the UK.

\subsection{Timeseries models for forecasting AI growth}

The expansion of the AI landscape in the UK is characterised by four entity lifecycle metrics: total AI entities, active AI entities, dissolved entities, and the entity dissolution ratio. The dataset spans the period from 2000 to 2024, with a designated forecasting horizon of 2025 to 2030. To project these trends, we employed four univariate time-series models: Auto-Regressive Integrated Moving Average (ARIMA), Theta, Exponential Smoothing (ETS), and Median Fourier Linear Exponential Smoothing (MFLES).

Hyperparameter optimisation was performed using the Tree-structured Parzen Estimator (TPE) algorithm \cite{bergstra2015hyperopt}. For this process, the data was split into a training set and a validation set (covering the period 2020–2024) to ensure robust parameter selection. Following optimisation, the models were re-trained on the complete time series using the identified optimal hyperparameters.

The predictive performance of the models was evaluated using Root Mean Squared Error (RMSE) (see eq. 2) and Mean Absolute Error (MAE). The mathematical formulation for MAE is given as:

\begin{equation}
MAE = \frac{\sum_{i=1}^N |y_i - \bar{y_i}|}{N}
\end{equation}

where $y_i$ represents the actual value, while $\hat{y}_i$ represents the predicted value. These metrics were computed for both training and validation phases to select the most effective forecasting model for each metric. The selected models were then deployed to forecast the AI lifecycle metrics for the 2025–2030 horizon. To account for uncertainty, we implemented Conformal Prediction to construct 95\% prediction intervals. This method was selected as it provides validated prediction intervals with marginal coverage guarantees, independent of strict distributional assumptions \cite{shafer2008tutorial}. 

\section{Results}
\subsection{The evolution of AI entities since 2000}

\begin{figure}[htp]
    \centering
    \includegraphics[width=1.0\linewidth]{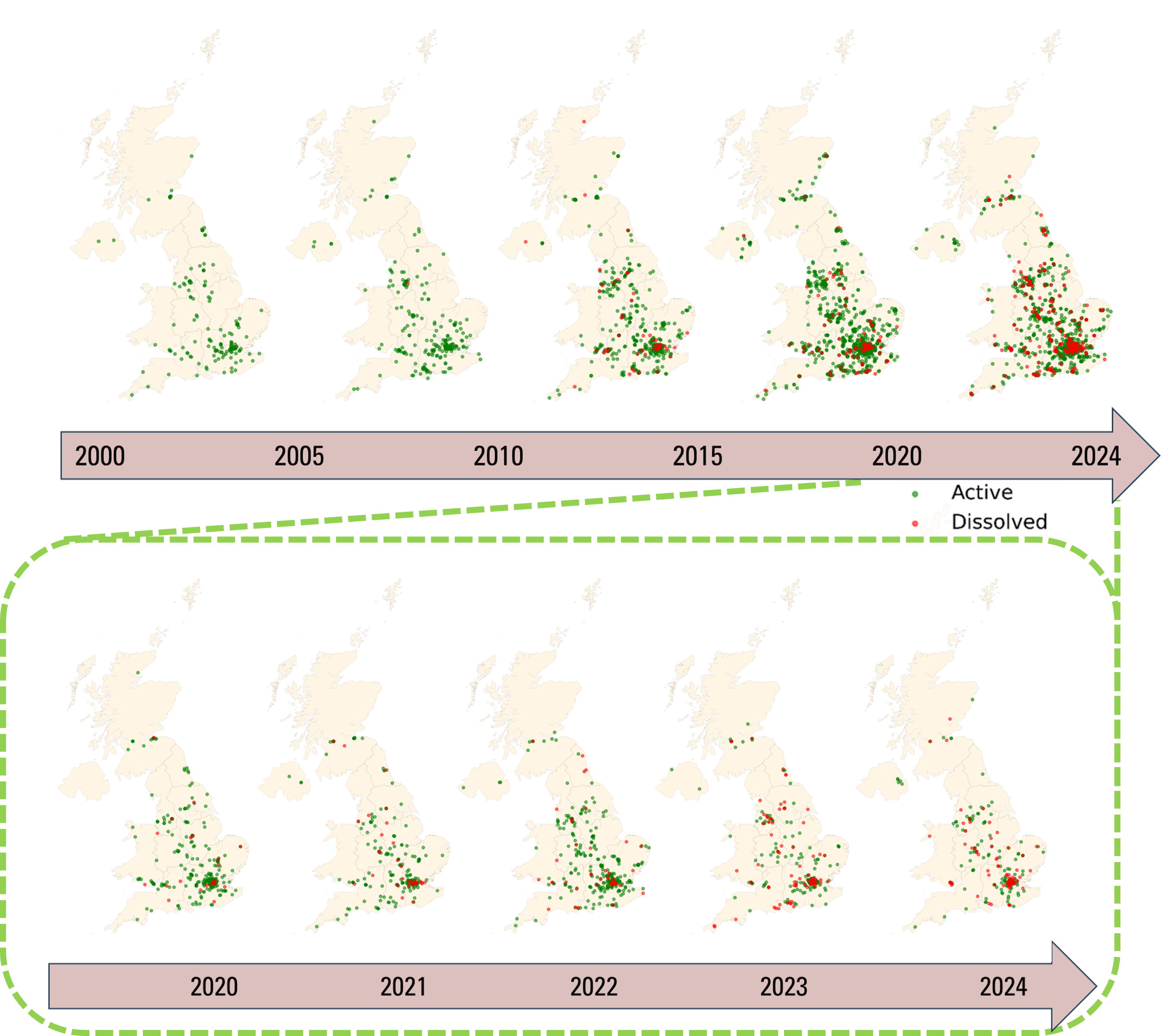}
    \caption{Evolution of AI Entities in the UK from 2000 to 2024. Variation in the count of AI entities corresponding to active and dissolved status is depicted with temporal-geographical details.}
    \label{Fig:fig_1}
\end{figure}

The evolution of AI entities in the UK from 2000 to 2024 is summarised in Figure~\ref{Fig:fig_1}. We report the key trends in registration and dissolution are presented in Table 1.

\begin{table}[htp]
    \centering
    \caption{Aggregate Trends in AI Entity Registration and Dissolution (2000--2024). \footnotesize{*Dissolved ratio calculated as (dissolved entities / new registrations) for the period.}}
    \begin{tabular}{lcccc}
        \toprule
        Period & New Registrations & \% Change on Previous Period & Dissolved Entities & Dissolved Ratio* \\
        \midrule
        2000--2005 & 181 & -- & 0 & 0.0\% \\
        2006--2010 & 267 & +47.5\% & 1 & 0.4\% \\
        2010--2015 & 592 & +121.7\% & 53 & 9.0\% \\
        2015--2020 & 1733 & +192.8\% & 110 & 6.3\% \\
        2020--2024 & 1607 & -7.3\% & 314 & 19.5\% \\
        \bottomrule
    \end{tabular}
    \smallskip
    \label{tab:aggregate_trends}
\end{table}

Our results for new registrations and the dissolved ratio relative to cumulative entities indicate strong momentum in the sector up to the onset of the COVID-19 pandemic  \cite{ralph2020pandemic}.  We estimated the dissolved ratio calculated as the dissolved entities per new registrations. In addition, we find that yearly registrations from 2020 to 2024 showed fluctuations, with new entities numbering 396, 309, 425, 258, and 219, respectively. Dissolutions in the same period were 27, 49, 50, 87, and 101, with the dissolved ratio increasing from 0.85\% to 2.29\% (refer to Figure~\ref{Fig:fig_5}).  Furthermore, the registration of AI entities from 2020 to 2024 was concentrated in England, with London accounting for 39.5\% (635) of the new entities. At the end of 2024, 41.3\% of all active AI entities (1,616) were based in London.  

\begin{figure}[htp]
    \centering
    \includegraphics[width=1.0\linewidth]{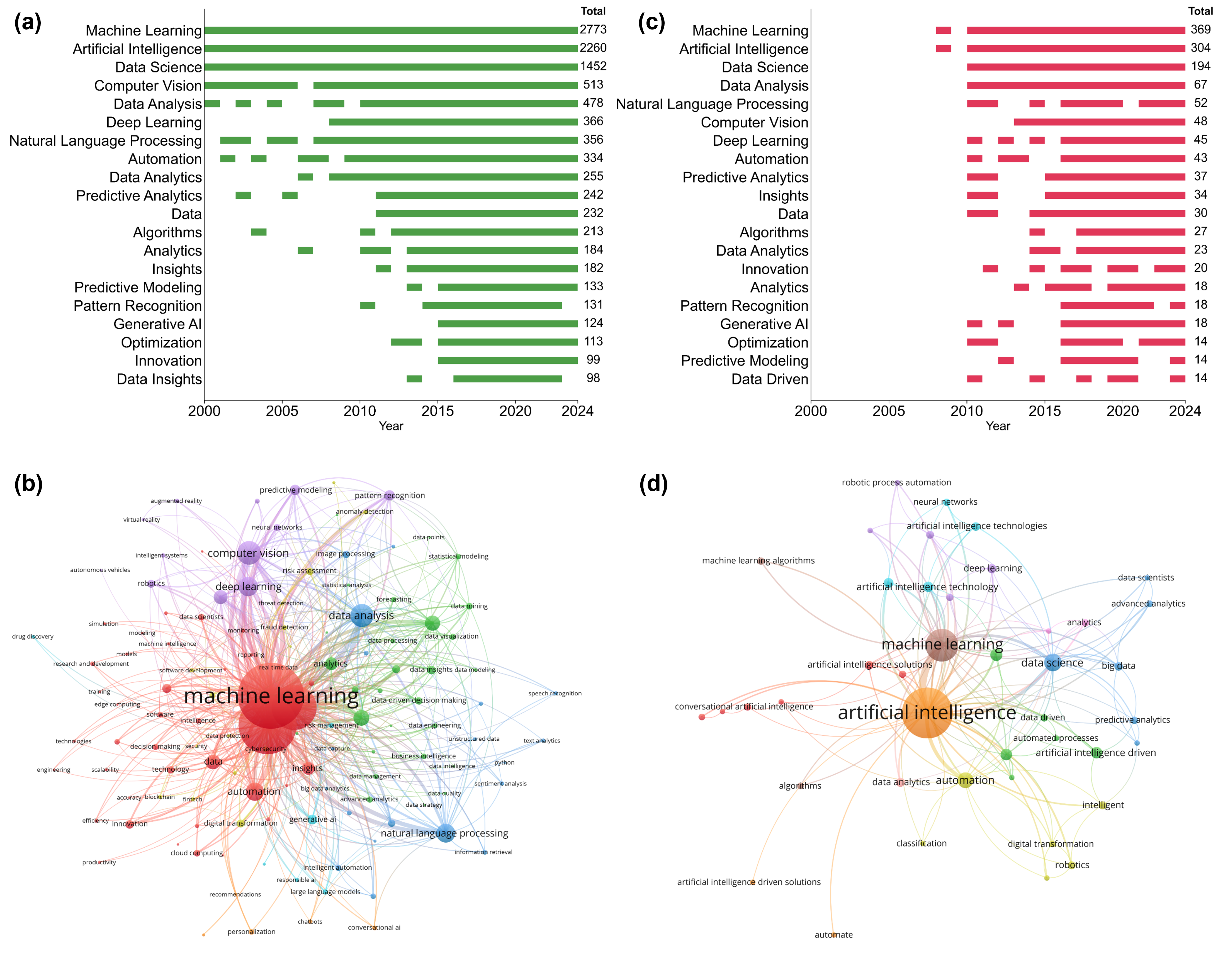}
    \caption{The evolution of keywords utilisation and their co-occurrences, respectively, (a) and (c) for Active, and (b) and (d) for Dissolved AI entities. }
    \label{Fig:fig_2}
\end{figure}

Figure~\ref{Fig:fig_2} shows the usage of keywords for active entities (n=3,143) and dissolved entities (n=398) from 2000--2010 (threshold frequency \(\geq\)3). For active entities, the most common keywords were \emph{machine learning} (2,773) and \emph{artificial intelligence} (2,260). These terms frequently co-occurred with others, indicating diverse service offerings. The use of \emph{generative AI} emerged around 2015 and remained prevalent in 2024, forming a distinct cluster with \emph{natural language processing} \cite{storey2025generative}. Keywords such as \emph{predictive modelling}, \emph{insights}, and \emph{optimisation} have been consistently used since approximately 2015.

Among dissolved entities, the keywords \emph{artificial intelligence} (369) and \emph{machine learning} (304) were also predominant. Keywords associated with current trends, including \emph{automation}, \emph{generative AI}, and \emph{insights}, were present among failed entities \cite{uriarte2025artificial}.

We find that in 2024, 29.5\% (141) of all dissolved entities were originally registered in London. A sectoral breakdown of London dissolutions shows Technology (80), Professional Services (13), Financial Services (11), and Healthcare/Scientific (11). Outside London, the Technology sector accounted for 198 dissolutions. Overall, 57.3\% of all dissolved entities were in the Technology sector. The median service years for dissolved entities in the UK were 4 years for Technology, Professional Services, and Financial Services. In London, the corresponding medians were 4, 3, and 1 years, respectively.

\subsection{Revenue generation by AI firms}

We compiled a detailed dataset for 529 AI entities, incorporating firm-level characteristics or features (employee count, service years, revenue, sector, SIC code, keyword score) and postcode-level socio-economic indicators. After excluding entities with incomplete records, the final analytical sample comprises 451 AI firms operating in the UK between 2000 and 2024.

\begin{figure}[htp]
    \centering
    \includegraphics[width=0.65\linewidth]{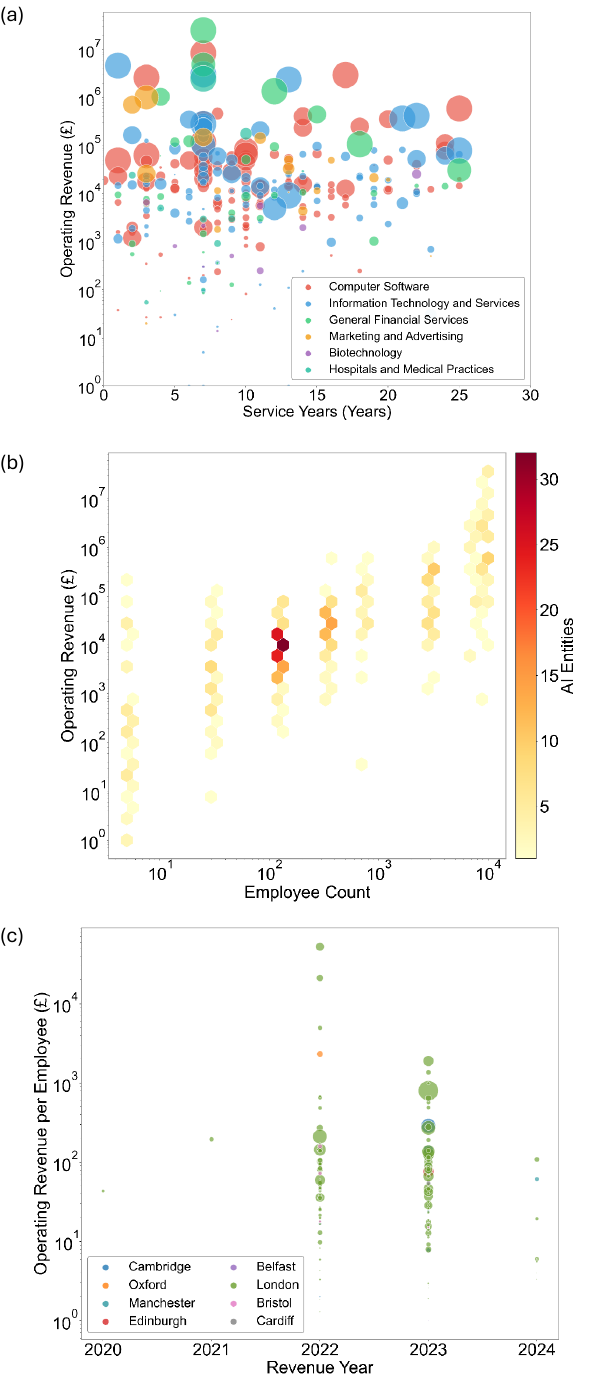}
    \caption{Association of revenue generated by AI entities with the firm-level variables and geographical footprint. (a) Scatter plot of operating revenue with service years, broken down with respect to sectors. (b) Binning the operating revenue against the employee count with respect to the similar count of AI entities. (c) The yearly dynamics of operating revenue per employee in the selected cities in the UK.}
    \label{Fig:fig_3}
\end{figure}

We show the relationship between operating revenue and firm maturity (service years) across key sectors in Figure~\ref{Fig:fig_3}(a). We find that revenue exhibits a growth phase during the first decade of operation for firms in computer software, information technology (IT) services, and general financial services, after which it tends to stabilise. The cumulative revenue generated by firms with at least 10 years of service varies considerably by sector (Table~\ref{tab:revenue_sector}). Across all sectors, the total lifetime revenue generated by the sampled firms amounts to approximately £284.7 million.

\begin{table}[htp]
    \centering
    \caption{Cumulative Revenue by Sector for AI Entities with $\geq10$ Service Years}
    \begin{tabular}{lcc}
        \toprule
        Sector & Cumulative Revenue (10+ years) & Sample Size (n) \\
        \midrule
        Computer Software & £13.2 million & 75 \\
        Information Technology \& Services & £10.7 million & 60 \\
        General Financial Services & £1.5 million & 21 \\
        Marketing \& Advertising & £1.9 million & 8 \\
        Biotechnology & £10.4 thousand & 6 \\
        Hospitals \& Medical Practices & £2.7 million & 9 \\
        \bottomrule
    \end{tabular}
    \label{tab:revenue_sector}
\end{table}

Additionally, figure~\ref{Fig:fig_3}(b) examines the association between operating revenue and employee count. While a positive correlation exists, the relationship is non-linear. Small firms ($\leq$10 employees) show high variability in revenue (range: £1 to £0.26 million; mean: £11.5 thousand), with some outperforming larger peers. Mid-sized firms (20–200 employees; mean employee count: 125) generate an average operating revenue of £13.5 thousand. The largest firms (>10,000 employees) achieve a mean revenue of £4.7 million, reflecting diversified business streams and economies of scale, as well as suggesting an accelerating growth trajectory for surviving and growing firms. This pattern also suggests that, while scale generally supports higher revenue, small, agile firms—often startups—can compete effectively in niche or high-growth segments.

A standard measure of firm success is operating revenue per employee, which serves as a proxy for labour productivity and can signal investment potential \cite{coyle2021uk}. Figure~\ref{Fig:fig_3}(c) tracks this metric over time for major UK cities. For example, we find that London consistently leads, with a peak of £52.4 thousand per employee in 2022—the highest value recorded in the 2020–2024 period.

\subsection{Drivers of AI firm performance}

To systematically assess the factors driving firm-level economic output, we developed two predictive models for operating revenue and operating revenue per employee, respectively. These models incorporate both firm-level characteristics—including keyword scores, service years, employee range, sector, and Standard Industrial Classification (SIC) codes—and postcode-level socio-economic variables such as population density, educational attainment (Level 3 / 4+), economic activity rates, and sectoral employment (e.g., finance, information and communication). The CatBoost algorithm, selected for its robustness with heterogeneous data, was trained and rigorously optimised via hyperparameter tuning.

\begin{figure}[htbp]
    \centering
    \includegraphics[width=1.0\linewidth]{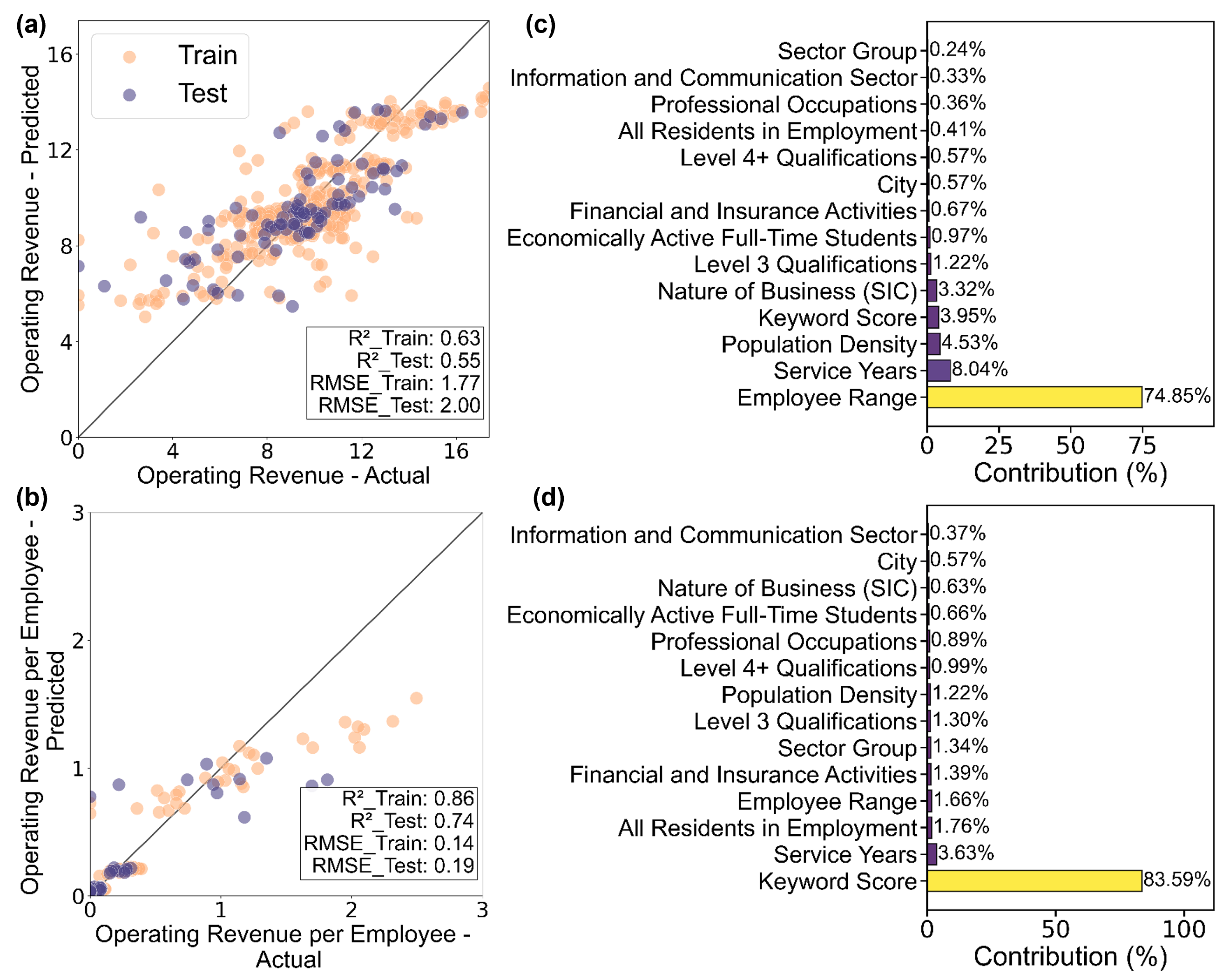}
    \caption{CatBoost-based modelling for (a) operating revenue and (b) operating revenue per employee of AI entities. SHAP framework provides the percentage contribution of input variable to predict (c) operating revenue and (d) operating revenue per employee for the trained models.}
    \label{Fig:fig_4}
\end{figure}

We evaluated the model performance on held-out test data. For operating revenue, the model achieved an \(R^2\) of 0.63 (RMSE: 1.77) on training and 0.55 (RMSE: 2.00) on testing data. For operating revenue per employee—a measure of labour productivity—performance was stronger, with an \(R^2\) of 0.86 (RMSE: 0.14) on training and 0.74 (RMSE: 0.19) on testing data (Fig.~\ref{Fig:fig_4}a,b). This indicates a robust ability to explain the variance in both outcome measures.

Furthermore, to interpret the models, we applied SHAP (Shapley Additive exPlanations) analysis, which quantifies the marginal contribution of each input variable to the predictions. The resulting relative importance scores reveal distinct drivers for the two economic metrics (Fig.~\ref{Fig:fig_4}c,d). For operating revenue, firm size (employee range) is the dominant predictor, accounting for 74.85\% of the explanatory power. Unsurprisingly larger firms tend to generate higher absolute revenues.

In contrast, operating revenue per employee is primarily explained by the firm's technological specialisation, as captured by the keyword score (83.59\% contribution). A higher keyword score reflects a broader or more relevant portfolio of AI capabilities—spanning established domains (e.g., computer vision, predictive analytics) and emerging ones (e.g., generative AI). This suggests that productivity is closely tied to a firm's ability to integrate and deploy relevant AI technologies, irrespective of its size. In particular, most firms exhibit keyword scores below 0.07, indicating a concentration in narrower technological niches and highlighting the potential for diversification.

We find that firm maturity (service years) also contributes positively to both outcomes (8.04\% for revenue; 3.63\% for revenue per employee), reflecting the benefits of accumulated market experience. However, the smaller contribution to productivity suggests that younger, technologically agile firms can also achieve high efficiency, consistent with observations of disruptive innovation in fast-evolving sectors \cite{kulkov2023next}. In addition, postcode-level socio-economic variables collectively account for meaningful variation in both models, with population density, employment rates, educational attainment (Level 3 qualifications) and local sectoral structure (e.g., financial and insurance activities) emerging as relevant contextual factors. This underscores the role of local talent pools and economic ecosystems in supporting AI firm performance.

These findings have two key implications. First, they highlight a dual-pathway to AI-led economic growth: scaling firm size drives aggregate revenue, while deepening technological specialisation enhances productivity. Second, the significant contribution of local socio-economic factors points to the importance of place-sensitive policies—such as investments in skill development and sector-specific infrastructure—to foster resilient AI ecosystems across UK regions.

\subsection{Forecasting AI firm growth in the UK}

We developed a set of time-series forecasts for key indicators of business formation and survival to project the future trajectory of the UK's AI sector. These indicators, the total number of registered AI entities, the count of currently active entities, the number dissolved annually, and the annual dissolution rate, collectively describe the sector's lifecycle dynamics. We compiled annual data for these four metrics from 2000 to 2024. The period 2020--2024 was held out as a validation set for model selection, while the full historical series (2000--2024) was used for final model training before generating forecasts through 2030.

We compared the predictive performance of four established time-series models: ARIMA (Auto Regressive Integrated Moving Average), ETS (Error, Trend, Seasonal), Theta, and MFLES (Multiple Fourier Linear Exponential Smoothing). The models were evaluated according to their Root Mean Square Error (RMSE) and Mean Absolute Error (MAE) in the training and validation datasets (Table~\ref{tab:performance_metrics}).

\begin{table}[h!]
    \centering
    \caption{Comparative performance of time-series forecasting models for AI sector lifecycle metrics. Lower MAE and RMSE values indicate better predictive accuracy. The best-performing model for each metric on the validation set is highlighted in bold.}
    \label{tab:performance_metrics}
    \begin{tabular}{llcccc}
        \toprule
        \multirow{2}{*}{\textbf{Forecast Variable}} & \multirow{2}{*}{\textbf{Model}} & \multicolumn{2}{c}{\textbf{Training (2000--2019)}} & \multicolumn{2}{c}{\textbf{Validation (2020--2024)}} \\
        \cmidrule(lr){3-4} \cmidrule(lr){5-6}
        & & \textbf{MAE} & \textbf{RMSE} & \textbf{MAE} & \textbf{RMSE} \\
        \midrule
        \multirow{4}{*}{Total AI Entities} & ARIMA & 36.18 & 51.31 & \textbf{25.67} & \textbf{30.01} \\
        & Theta & 82.33 & 115.11 & 304.45 & 326.04 \\
        & ETS & 83.24 & 119.56 & 31.12 & 34.26 \\
        & MFLES & 124.31 & 157.05 & 49.99 & 54.04 \\
        \midrule
        \multirow{4}{*}{Active AI Entities} & ARIMA & 37.73 & 56.88 & 45.95 & 49.83 \\
        & Theta & 78.79 & 110.10 & 190.11 & 203.93 \\
        & ETS & 69.89 & 100.46 & \textbf{42.95} & \textbf{46.80} \\
        & MFLES & 64.88 & 76.80 & 62.57 & 71.76 \\
        \midrule
        \multirow{4}{*}{Dissolved AI Entities} & ARIMA & 3.41 & 5.61 & 32.01 & 37.47 \\
        & Theta & 3.41 & 5.86 & 38.55 & 44.03 \\
        & ETS & 4.21 & 6.29 & 23.24 & 28.72 \\
        & MFLES & 4.16 & 7.09 & \textbf{10.34} & \textbf{11.10} \\
        \midrule
        \multirow{4}{*}{Dissolution Rate} & ARIMA & 0.33 & 0.46 & \textbf{0.21} & \textbf{0.27} \\
        & Theta & 0.30 & 0.70 & 0.62 & 0.74 \\
        & ETS & 0.37 & 0.77 & 0.29 & 0.31 \\
        & MFLES & 0.57 & 0.97 & 0.18 & 0.21 \\
        \bottomrule
    \end{tabular}
\end{table}

We find that performance analysis reveals that no single model dominates in all metrics, necessitating a specific metric selection strategy. The ARIMA model demonstrated the strongest predictive accuracy for \textit{Total AI Entities} and the \textit{Dissolution Rate} (annual dissolved entities as a proportion of the active stock). The ETS model was selected for forecasting \textit{Active AI Entities}, while the MFLES model performed best for \textit{Dissolved AI Entities}. These selected models were used to generate the final forecasts presented in Figure~\ref{Fig:fig_5}.

The forecasts point to a maturing AI sector in the United Kingdom with distinct but linked dynamics for entry, activity, and exit (Figure~\ref{Fig:fig_5}). We infer this pattern as follows:

\begin{figure}[htbp]
    \centering
    \includegraphics[width=1.0\linewidth]{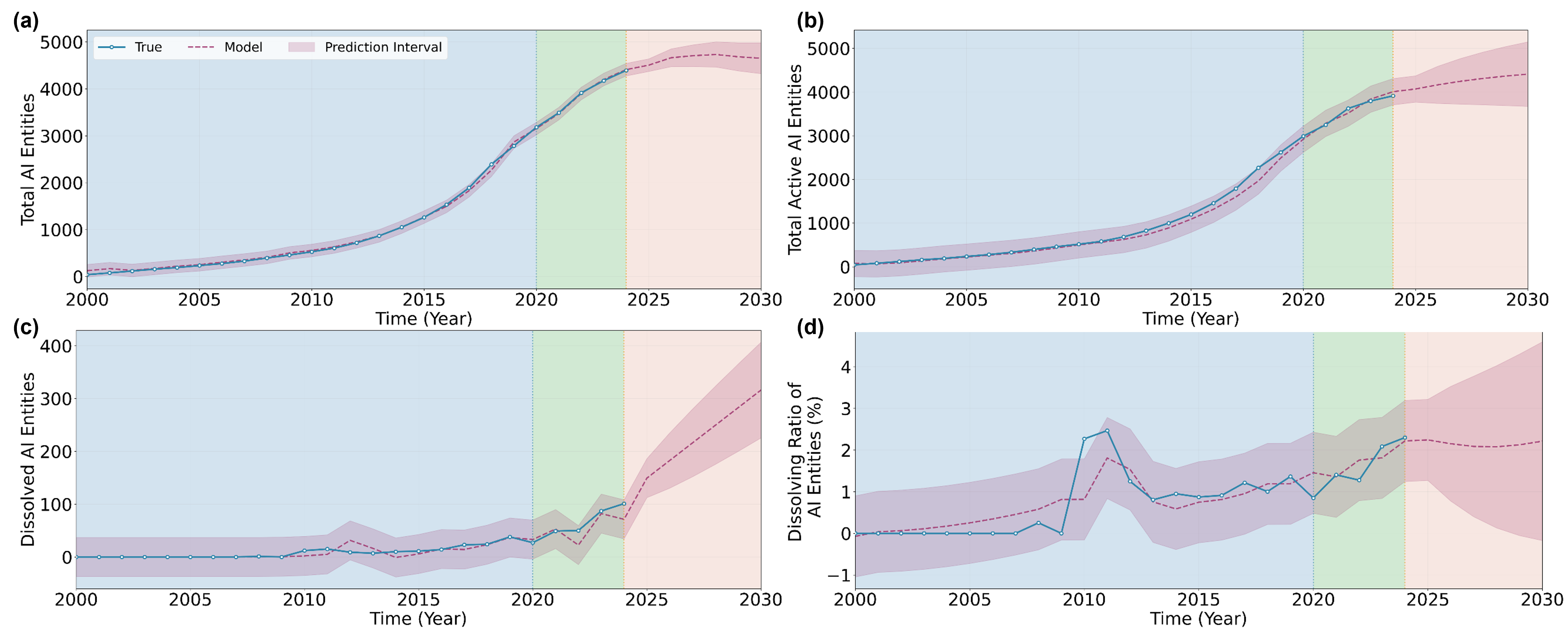}
    \caption{Historical trends and model forecasts for AI sector dynamics in the UK, 2000--2030. (a) Total registered AI entities. (b) Active AI entities. (c) Annual dissolutions of AI entities. (d) Annual dissolution rate (dissolutions/active entities). The blue series represents the training data (2000--2019), the green series the validation period (2020--2024), and the orange series the model forecast with 95\% prediction interval (2025--2030). The dashed vertical lines separate these periods.}
    \label{Fig:fig_5}
\end{figure}

\textbf{Sector growth is projected to plateau:} The historical accumulation of registered AI entities grew exponentially after 2010, surpassing 3,000 by 2020. However, the growth momentum has slowed considerably after 2020. The ARIMA forecast suggests that this trend will continue, projecting a near-flat trajectory for the total stock of entities, reaching approximately 4,651 (95\% CI: 4,323--4,979) by 2030. This indicates a potential saturation of new market entries.

\textbf{The active enterprise base will continue modest growth:} The count of active entities as a more direct indicator of economic activity showed strong growth until 2020, stabilising thereafter at around 3,900 in 2024. The ETS model forecasts a resumption of moderate and stable growth in the active base, projecting it to reach 4,410 (95\% PI: 3,672--5,149) by 2030. The widening prediction interval in later years reflects increasing uncertainty.

\textbf{Business dissolutions are forecast to rise:} The dissolution of AI entities, which remained negligible until after the 2008 financial crisis, has exhibited a clear upward trend since 2013, reaching 101 dissolutions (a 2.29\% annual rate) in 2024. The MFLES model forecasts this count to rise significantly to 316 (95\% PI: 226--407) by 2030. Concurrently, the ARIMA model forecasts the annual dissolution rate to remain stable at approximately 2.21\% (95\% PI: -0.17\%--4.60\%). The stable rate alongside rising absolute numbers is consistent with a growing but consolidating sector where the annual failure rate normalises even as the larger pool of firms yields more exits.

Collectively, these forecasts describe a sector that is transitioning from a high-growth phase to a more mature consolidation phase. The plateau in new registrations, coupled with rising absolute dissolutions, suggests a competitive landscape where market entry is becoming more challenging and less productive firms are exiting. However, continued modest growth in the active base implies resilience and sustained economic activity within the core of the sector.

\section{Discussion and Conclusion}

Our empirical analysis reveals three critical dynamics shaping the United Kingdom's artificial intelligence ecosystem: pronounced geographical concentration in London, productivity drivers tied to firm scale and technical specialization, and an emerging phase of market consolidation. These findings, derived from comprehensive firm-level and postcode-linked data, provide evidence-based insights with significant implications for the UK's ambition to become a science and technology superpower.

The London-centric distribution of AI entities, with capital accounting for more than 40\% of registered firms and demonstrating the highest productivity, echoes patterns observed in other knowledge-intensive sectors \cite{dsit2025sector}. This concentration reflects agglomeration economies, but also presents systemic risks, including regional inequality and reduced national economic resilience. Our identification of Level 3 qualifications, population density, and local employment rates as significant postcode-level correlates with firm revenue underscores that AI development is not geographically neutral; it is embedded within and dependent upon local socio-economic ecosystems. This finding challenges purely sectoral innovation policies and suggests that place-sensitive interventions may be required to catalyse growth beyond existing hubs.

In this study, machine learning analysis identifies the range of employees and keyword specialisation scores as the primary determinants of operating revenue and revenue per employee, respectively. This indicates that the UK's AI sector faces distinct challenges at different lifecycle stages: initial scaling (increasing employee base) and subsequent deepening of technical specialisation to drive productivity. The prominence of financial services, computer software, and IT services in revenue generation suggests that application-orientated AI currently dominates commercial success, potentially at the expense of frontier research commercialisation.

Our forecasting models project a deceleration in AI firm growth along with a stable but rising dissolution rate by 2030 (aligned with the UK AI 2030 scenarios \cite{ukgov2024AIproject}). This pattern resembles the maturation phase observed in other technology sectors \cite{nguyen2025start}, where market consolidation follows initial explosive growth. Rather than indicating sectoral decline, this trajectory may signal a transition toward a more stable, quality-focused ecosystem. However, without strategic intervention, this consolidation risks reinforcing existing geographical and sectoral imbalances.

Several limitations constrain the scope and generalisability of our findings, highlighting areas for refinement in understanding the UK's AI landscape. First, our analysis is predicated on company registration and financial reporting datasets, which can underrepresent very early-stage startups or firms that do not fully disclose financial information. This potential bias toward more established entities could obscure the dynamics of nascent innovation, thereby skewing insights into overall sector growth and productivity drivers. Second, the classification of AI specialisations through keyword scores effectively captures broad technological focuses, but falls short in addressing nuanced differences, such as variations in product maturity, intellectual property portfolios, or sector-specific applications. This high-level approach may oversimplify the complex interplay between technical specialisation and firm performance, limiting the depth of our postcode-level socio-economic correlations and revenue generation analyses. Third, while our study emphasises the UK's context, the lack of direct comparisons with other leading AI ecosystems (e.g., the US, China, and the EU) restricts our ability to contextualise the role of technical specialisation and socio-economic characteristics relative to global benchmarks. This omission could overlook external influences, such as international funding flows or regulatory environments, that shape the UK's AI evolution.

Future research should address these gaps by diversifying data sources—such as incorporating venture capital records, patent databases, or qualitative surveys—to better capture early-stage activities and finer-grained firm attributes. Additionally, in-depth case studies of regional AI clusters, longitudinal tracking of firm trajectories, and cross-national comparative analyses would provide a more robust framework for evaluating policy interventions and fostering a balanced AI ecosystem in the UK.

However, the evidence presented here supports a reorientation of UK AI policy from general support toward targeted interventions that address three strategic imperatives.

\textbf{First, policy must actively cultivate regional AI capabilities and ecosystems,} because current market forces alone seem unlikely to alter the geographical concentration. We consider there is scope for establishing at least three additional AI superclusters outside London (such as Cambridge, Manchester and Bristol), supported by tailored combinations of research infrastructure (e.g., regional compute resources), skills development (expanding AI conversion courses regionally), and connectivity to local industry strengths. This approach recognises that successful AI development depends on complementary assets that vary by region.

\textbf{Second, policy should distinguish between scaling and specialisation challenges.} For scaling, addressing the UK's well-documented growth capital gap through sector-specific funds and public procurement reform is essential. For specialisation, creating clear pathways from academic research to commercialisation—particularly in areas of UK scientific strength like life sciences and climate—could enhance productivity. The AI regulatory framework should balance risk mitigation with support for innovative applications in these sectors.

\textbf{Third, policy must prepare for and shape ecosystem consolidation:} Rather than focus solely on startup creation, support mechanisms should help viable firms navigate growth transitions and facilitate knowledge-preserving exits when dissolution occurs. Monitoring dissolution patterns will provide early warning signals of sectoral stress.

The UK is at a pivotal moment in its AI development, given the slower growth projected on the basis of our evidence. Strategic, evidence-based policy intervention can steer the UK toward a more geographically balanced, productive, and resilient ecosystem—transforming what could be a period of consolidation into an opportunity to build a lasting competitive advantage in artificial intelligence.

\section*{Acknowledgment}
RD acknowledges the funding received from UKRI Responsible AI IO0008 Grant Ref: EP/Y009800/1 and support from ai@cam through the AIDEAS Grant. The authors also acknowledge the support received from Glass.ai and Sergi Martorell who shared the data for this study. The authors thank the wider RAI UK project team for valuable comments during the team workshops.

\section*{Data and Code Availability}
We established a baseline using the UK Research and Innovation (UKRI) "WAIFinder" database, which catalogues 3,280 AI-related organisations \cite{ukri2024waifinder}. This baseline was subsequently updated and expanded to cover the period up to 2024 with the support of glass.ai and data taken from Company House \cite{companieshouse2025} and Office of National Statistics \cite{ons2025census}. The code made for the analysis is available at https://github.com/Waqar9871/RAI\_UK.git. The data from glass.ai are proprietary information of glass.ai.

\printbibliography

\end{document}